# MIMIC modelling with instrumental variables: A 2SLS-MIMIC approach

Andrej Srakar, Institute for Economic Research (IER) and School of Economics and Business, University of Ljubljana, Slovenia, srakara@ier.si

Marilena Vecco, Burgundy School of Business – Université Bourgogne Franche-Comté, France, marilena.vecco@bsb-education.com

Miroslav Verbič, School of Economics and Business, University of Ljubljana and Institute for Economic Research (IER), Ljubljana, Slovenia, miroslav.verbic@ef.uni-lj.si

Montserrat Gonzalez Garibay, Institute for Economic Research (IER), gonzalezm@ier.si

Jože Sambt, School of Economics and Business, University of Ljubljana, Slovenia, joze.sambt@ef.uni-lj.si

## Abstract

Multiple Indicators Multiple Causes (MIMIC) models are type of structural equation models, a theory-based approach to confirm the influence of a set of exogenous causal variables on the latent variable, and also the effect of the latent variable on observed indicator variables. In a common MIMIC model, multiple indicators reflect the underlying latent variables/factors, and the multiple causes (observed predictors) affect latent variables/factors. Basic assumptions of MIMIC are clearly violated in case of a variable being both an indicator and a cause, i.e. in the presence of reverse causality. Furthermore, the model is then unidentified. To resolve the situation, which can arise frequently, and as MIMIC estimation lacks closed form solutions for parameters we utilize a version of Bollen's (1996) 2SLS estimator for structural equation models combined with Jöreskog (1970)'s method of the analysis of covariance structures to derive a new, 2SLS estimator for MIMIC models. Our 2SLS empirical estimation is based on static MIMIC specification but we point also to dynamic/error-correction MIMIC specification and 2SLS solution for it. We derive basic asymptotic theory for static 2SLS-MIMIC, present a simulation study and apply findings to an interesting empirical case of estimating precarious status of older workers (using dataset of Survey of Health, Ageing and Retirement in Europe) which solves an important issue of the definition of precarious work as a multidimensional concept, not modelled adequately so far.

## 1. Introduction

Multiple Indicators Multiple Causes (MIMIC) models are type of structural equation models, a theory-based approach to confirm the influence of a set of exogenous causal variables on the latent variable, and also the effect of the latent variable on observed indicator variables (see e.g. Zellner, 1970; Goldberger, 1972; Jöreskog and Goldberger, 1975; Weck, 1983; Frey and Weck, 1983; Frey and Weck-Hannemann, 1984; Aigner et al., 1988; for some more recent applications see e.g. Lester, 2008; Proitsi et al., 2009; Rose and Spiegel, 2010). MIMIC models are commonly used in economics for modelling the underground economy (for discussion on this topic see e.g. Thomas, 1992; Schneider, 1994; 1997; 2003; 2005; Lippert and Walker, 1997; Johnson et al., 1998a, 1998b; Tanzi, 1999; Giles, 1999; Mummert and Schneider, 2001; Giles and Tedds, 2002; Giles et al., 2002; Dell'Anno and Schneider, 2003; Buehn and Schneider, 2008; Barbosa et al., 2013; Nchor and Adamec, 2015; Breusch, 2016).

In a common MIMIC model, multiple indicators reflect the underlying latent variables/factors, and the multiple causes (observed predictors) affect latent variables/factors. Basic assumptions of MIMIC are clearly violated in case of a variable being both an indicator and a cause, i.e. in the presence of reverse causality. Furthermore, the model is then unidentified. To resolve the situation, which can arise frequently (for example, in modelling shadow economy, GDP can be both a predictor and consequence), we utilize a version of Bollen's (1996) 2SLS estimator for structural equation models combined with Jöreskog (1970)'s method of the analysis of covariance structures to derive a new, 2SLS estimator for MIMIC models. As MIMIC estimation lacks closed form solutions for parameters we study and use Madansky-Hägglund-Jöreskog and Bollen's IV and 2SLS approaches to estimate the covariance matrices of latent variables. Second, we use this estimated covariance matrix of the latent variables and apply Jöreskog's (1970) maximum likelihood procedure to estimate coefficient estimates for the latent variable model. Our 2SLS empirical estimation is based on static MIMIC specification but we point also to dynamic/error-correction (Buehn and Schneider, 2008) MIMIC specification and 2SLS solution for it. We derive basic asymptotic theory for static 2SLS-MIMIC, present a simulation study and apply findings to an interesting empirical case of estimating precarious status of older workers (using dataset of Survey of Health, Ageing and Retirement in Europe) which solves an important issue of the definition of precarious work as a multidimensional concept, not modelled adequately so far.

## 2. 2SLS-MIMIC derivation and properties

The formal mathematical representation of the MIMIC model reads as follows (see e.g. Hodge and Traiman, 1968; Jöreskog and Goldberger, 1975):

$$y^* = \alpha' x + \epsilon \quad (1)$$
$$y = \beta y^* + u \quad (2)$$

where $y = (y_1, y_2, \ldots, y_p)'$ are indicators of the latent variable $y^*$ and $x = (x_1, x_2, \ldots, x_q)'$ are causes of $y^*$.

The model is based on the following assumptions (see e.g. Jöreskog and Goldberger, 1975; Trebicka, 2014):

$$E(\epsilon u') = 0', E(\epsilon^2) = \sigma^2, E(uu') = \Theta^2 \quad (3)$$

Applications of MIMIC models, typically, also make distributional assumptions, for example that the joint distribution of the variables is Gaussian, the relation is linear, and each measured variable and each latent common cause has specific sources of variance that are independent of the sources of variance specific to other variables (Trebicka, 2014).

Clearly, if $x$ and $y$ are in endogeneous relationship, for example if both have an influence to each other (so-called reverse causality) assumptions in (3) are violated. In this case it is also impossible to identify the relationships in (1) and (2).

Deriving from (1) and (2), the reduced form model and formula for variance-covariance matrix is described in (Jöreskog and Goldberger, 1975):

$$y = \beta(\alpha'x + \epsilon) + u = \Pi'x + v \qquad (4)$$
$$\Pi = \alpha\beta' \qquad (5)$$
$$v = \beta\epsilon + u \qquad (6)$$
$$\Omega = E(vv') = E[(\beta\epsilon + u)(\beta\epsilon + u)'] = \sigma^2\beta\beta' + \Theta^2 \qquad (7)$$

The formulas for the MIMIC parameters ($\alpha$, $\beta$, $\Theta$) cannot be expressed in closed form (Jöreskog and Goldberger, 1975). Implicit forms can be derived following Jöreskog and Goldberger (1975) as:

$$\hat{\alpha} = \left(\frac{1}{\hat{\kappa}^2}\right)P\hat{\Omega}^{-1}\hat{\beta} = \left(\frac{1}{\hat{\pi}^2}\right)\hat{P}\Theta^{-2}\hat{\beta} \qquad (8)$$

$$\left[S + \left(\frac{1}{\hat{\kappa}^2}\right)Q\right]\hat{\Omega}^{-1}\hat{\beta} = (1 + \hat{\rho}^2)\hat{\beta} \qquad (9)$$

$$\pi^2 = \beta'\Theta^{-2}\beta, \kappa^2 = \beta'\Omega^{-1}\beta = \frac{\pi^2}{(1+\pi^2)} \qquad (10)$$

$$P = (X'P_XX)^{-1}X'P_XY, Q = Y'XP \qquad (11)$$

$$S = (Y - XP_XP)')(Y - XP_XP) = Y'(I - XP_X(X'P_XX)^{-1}P_XX')Y \qquad (12)$$

$$P_X = X(X'X)^{-1}X' \qquad (13)$$

To derive the properties of a new estimator able to correct for the violation of the assumptions in (3) due to endogeneity in the model, we use Jöreskog's (1970) method of the analysis of covariance structures and proposal from Jöreskog and Goldberger (1975) to transform the MIMIC model into Jöreskog (1970) covariance structure modelling framework.

Jöreskog (1970) develops a general covariance structure model for a multivariate normal vector z as:

$$E(z'z) = \Sigma = B(\Lambda\Phi\Lambda' + \Psi^2)B' + \Theta^2 \qquad (14)$$
$$E(z) = A\Xi P \qquad (15)$$

where A is an N × g matrix of rank g and P is a h × p matrix of rank h, both being fixed matrices with g ≤ N and h ≤ p; $\Xi$, B, $\Lambda$, the symmetric matrix $\Phi$, and the diagonal matrices $\Psi$ and $\Theta$ are parameter matrices.

Based on the model in (14) and (15), Jöreskog derives the log-likelihood function as:

$$\log L = -\frac{1}{2} pN \log(2\pi) - \frac{1}{2} N \log|\Sigma| - \frac{1}{2} \sum_{a=1}^{N} \sum_{i=1}^{p} \sum_{j=1}^{p} (x_{ai} - \mu_{ai}) \sigma^{ij} (x_{aj} - \mu_{aj}) \qquad (16)$$

where $\mu_{ai}$ and $\sigma^{ij}$ are elements of $E(X) = A\Xi P$ and $\Sigma^{-1}$, respectively.

Writing

$$T = \frac{1}{N} (X - A\Xi P)'(X - A\Xi P) \qquad (17)$$

we can readily see that maximizing $\log L$ is equivalent to minimizing

$$F = \log|\Sigma| + tr(T\Sigma^{-1}) \qquad (18)$$

For MIMIC model, taking $z = (x', y')$ we have in the random case

$$\Sigma = \begin{pmatrix} \Phi & \Phi\alpha\beta' \\ \beta\alpha'\Phi & (1 + \rho^2)\beta\beta' + \Theta^2 \end{pmatrix} \qquad (19)$$

This covariance structure may be specified in terms of Jöreskog's model by setting

$$B = \begin{pmatrix} I_{k \times k} & 0_{k \times 1} \\ 0_{m \times k} & \beta_{m \times 1} \end{pmatrix}, \Lambda = \begin{pmatrix} I_{k \times k} \\ \alpha'_{1 \times k} \end{pmatrix},$$
$$\Psi = \begin{pmatrix} 0_{k \times k} & 0_{k \times 1} \\ 0_{1 \times k} & I_{1 \times 1} \end{pmatrix}, \Theta = \begin{pmatrix} 0_{k \times k} & 0_{k \times m} \\ 0_{m \times k} & \Theta_{m \times m} \end{pmatrix} \qquad (20)$$

and taking $\Phi$ free (Jöreskog and Goldberger, 1975).

As our parameters and estimator cannot be expressed in closed form, we adopt a strategy from Jöreskog and Sorbom (1993), who use 2SLS estimator for equations from the latent variable model using two part strategy: first, using Madansky-Hägglund-Jöreskog and Bollen's IV and 2SLS estimators for the factor loadings of the measurement model along with formulas from Hägglund (1982) to estimate the covariance matrices of latent variables. Second, we use this estimated covariance matrix of the latent variables and apply Jöreskog's (1970) maximum likelihood procedure to estimate coefficient estimates for the latent variable model. Our analysis was firstly presented in Srakar, Vecco and Verbič (present version 2020) and applied in Vecco and Srakar (2018).

In his seminal article, Bollen (1996) constructs a new 2SLS estimator for structural equation models, based on limited information maximum likelihood, as follows. His initial latent variable model reads as:

$$\eta = \alpha + B\eta + \Gamma\xi + \zeta \qquad (21)$$

where $\eta$ is an $m \times 1$ vector of latent endogenous random variables, $B$ is a $m \times m$ matrix of coefficients that give the impact of the $\eta$'s on each other, $\xi$ is an $n \times 1$ vector of latent exogenous variables, $\Gamma$ is the $m \times n$ coefficient matrix giving $\xi$'s impact on $\eta$, $\alpha$ is an $m \times 1$ vector of intercept terms, and $\zeta$ is an $m \times 1$ vector of random disturbances with the $\mathbb{E}(\zeta) = 0$ and $Cov(\xi, \zeta') = 0$.

Writing $y_1 = \eta + \varepsilon_1$ and $x_1 = \xi + \delta_1$ above equation transforms into:

$$y_1 = \alpha + By_1 + \Gamma x_1 + u \qquad (22)$$

where $u = \varepsilon_1 - B\varepsilon_1 - \Gamma\delta_1 + \zeta$.

Bollen considers a single equation from $y_1$ as:

$$y_i = \alpha_i + B_i y_1 + \Gamma_i x_1 + u_i \qquad (23)$$

where $y_i$ is the $i$-th $y$ from $y_1$, $\alpha_i$ is the corresponding intercept, $B_i$ is the $i$-th row from $By_1$, $\Gamma_i$ is the $i$-th row from $\Gamma$, and $u_i$ is the $i$-th element from $u$.

Defining $A_i$ to be a column vector that contains $\alpha_i$ and all the nonzero elements of $B_i$ and $\Gamma_i$ strung together in a column. Let $N$ equal the number of cases and $Z_i$ be an $N$ row matrix that contains 1's in the first column and the $N$ rows of elements from $y_1$ and $x_1$ that have nonzero coefficients associated with them in the remaining columns. The $N \times 1$ vector $y_i$ contains the $N$ values of $y_i$ contains the $N$ values of $y_i$ in the sample and $u_i$ is an $N \times 1$ vector of the values of $u_i$. The we can rewrite above as:

$$y_i = Z_i A_i + u_i \qquad (24)$$

As ordinary least squares is inappropriate for this estimation, we can use two-stage least squares (2SLS) estimator as an alternative and consistent estimator of $A_i$.

The 2SLS estimator require instrumental variables for $Z_i$. They must be: a) correlated with $Z_i$, b) uncorrelated with $u_i$, and c) sufficient in number so that there at least as many IV's as the number of explanatory variables on the right-hand side of the equation. Generally, in Bollen's 2SLS estimation, the pool of potential IVs comes from those $y$'s and $x$'s not included in $Z_i$ (excluding, of course, $y_i$). The exceptions are any variables in $Z_i$ that are uncorrelated with $u_i$, since such variables can serve as IVs. Exogenous (predetermined) $x$'s would be an example of IVs that might appear on the right-hand side above.

Assume we collect all eligible IVs for $Z_i$ and a column of 1's in an $N$ row matrix $V_i$. Then the first stage of 2SLS is to regress $Z_i$ on $V_i$ where below provides the coefficient estimator:

$$(V_i'V_i)^{-1}V_i'Z_i \qquad (25)$$

Form $\hat{Z}_i$ as:

$$\hat{Z}_i = V_i(V_i'V_i)^{-1}V_i'Z_i \qquad (26)$$

The second stage is the OLS regression of $y_i$ on $\hat{Z}_i$ so that

$$\hat{A} = \left(\hat{Z}_i'\hat{Z}_i\right)^{-1}\hat{Z}_i'y_i \qquad (27)$$

It can be easily shown (Bollen, 1996) that the estimator is consistent and asymptotically normal, as holds for the usual 2SLS estimators from econometrics.

Following Buehn and Schneider (2008), we can write the variance-covariance matrix of the 2SLS-MIMIC as:

$$\Sigma^* = \begin{pmatrix} \Phi^* & \Phi^* \alpha^* \beta' \\ \beta \alpha^{*\prime} \Phi^* & (1+\rho^{*2})\beta\beta' + \Theta^2 \end{pmatrix} \quad (28)$$

$$\Phi^* = X'P_X X \quad (29)$$

$$\rho^* = \alpha^{*\prime} X' P_X X \alpha^* \quad (30)$$

$$\hat{\alpha}^* = \left(\frac{1}{\hat{\kappa}^2}\right)(X'P_X X)^{-1} X' P_X Y \hat{\Omega}^{-1} \hat{\beta} \quad (31)$$

$$P_X = X(X'X)^{-1}X' \quad (32)$$

This covariance structure may be specified in terms of Jöreskog's model by setting

$$B = \begin{pmatrix} I_{k \times k} & 0_{k \times 1} \\ 0_{m \times k} & \beta_{m \times 1} \end{pmatrix}, \Lambda = \begin{pmatrix} I_{k \times k} \\ \alpha^{*\prime}_{1 \times k} \end{pmatrix},$$
$$\Psi = \begin{pmatrix} 0_{k \times k} & 0_{k \times 1} \\ 0_{1 \times k} & I_{1 \times 1} \end{pmatrix}, \Theta = \begin{pmatrix} 0_{k \times k} & 0_{k \times m} \\ 0_{m \times k} & \Theta_{m \times m} \end{pmatrix} \quad (33)$$

and taking $\Phi^*$ free.

The final parameter estimation is performed following Jöreskog and Goldberger (1975)'s suggestion of using maximum likelihood estimation, firstly transforming the above problem into Jöreskog (1970)'s covariance structure modelling framework.

The performance of the new estimator can be based on previous findings. Let $F$ denote a maximum likelihood objective function, $\eta$ the estimated moment structure and $\vartheta$ vector of parameters. We can then write the Hessian matrices as:

$$H_{\eta\vartheta} = \left.\frac{\partial^2 F(u, \eta(\vartheta))}{\partial u \partial \vartheta'}\right|_{u=\eta_0, \vartheta=\vartheta_0} \quad (34)$$

$$H_{\vartheta\vartheta} = \left.\frac{\partial^2 F(u, \eta(\vartheta))}{\partial \vartheta \partial \vartheta'}\right|_{\vartheta=\vartheta_0} \quad (35)$$

Under regularity assumptions it may be shown (Dijkstra, 1983; Shapiro, 1983) that the limiting distribution of $N^{1/2}(\hat{\vartheta} - \vartheta_0)$ as $N \to \infty$ is multivariate normal with a null mean and a covariance matrix

$$\Pi = H_{\vartheta\vartheta}^{-1} H'_{\eta\vartheta} \Gamma_0 H_{\eta\vartheta} H_{\vartheta\vartheta}^{-1} \quad (36)$$

where

$$\Gamma_0 = \begin{bmatrix} \Sigma_0 & \Gamma_{0_{\vartheta\omega}} \\ \Gamma_{0_{\omega\vartheta}} & \Gamma_{0_{\omega\omega}} \end{bmatrix} \quad (37)$$

with

$$\Gamma_{0_{\vartheta\omega}} = \mathbb{E}[(y - \mu_0) vecs\{(y - \mu_0)(y - \mu_0)'\}] \quad (38)$$

with typical element

$$[\Gamma_{0_{\vartheta\omega}}]_{i,jk} = \mathbb{E}(y_i - \mu_{0i})(y_j - \mu_{0j})(y_k - \mu_{0k}) \qquad (39)$$

and

$$\Gamma_{0_{\omega\omega}} = \mathbb{E}[vecs\{(y - \mu_0)(y - \mu_0)'\}vecs\{(y - \mu_0)(y - \mu_0)'\}] - \sigma_0\sigma_0' \qquad (40)$$

with typical element

$$[\Gamma_{0_{\omega\omega}}]_{ij,kl} = \mathbb{E}(y_i - \mu_{0i})(y_j - \mu_{0j})(y_k - \mu_{0k})(y_l - \mu_{0l}) - \sigma_{0ij}\sigma_{0kl} \qquad (41)$$

so that $\Gamma_0$ depends on second, third and fourth order central moments of the distribution of $y$.

As the data entering 2SLS-MIMIC estimation are independent and identically distributed, we can apply reasoning of Lee and Shi (1998) and argue that the derived estimator is consistent, asymptotically normal and efficient. This shows the consistency of our procedure and main properties of the derived estimation process which guarantee the desired behavior of the estimates.

### 3. 2SLS-EMIMIC derivation and simulation study

In MIMIC models, it is also possible to consider a dynamic situation. Often, MIMIC models are applied to time series data to derive, for example, estimates of the size and development of the shadow economy over time. As most macroeconomic variables do not satisfy the underlying assumption of stationarity, the problem of spurious regressions may arise. Researchers usually overcome this problem by transforming the time series into stationary ones, employing a difference operator. Alternatively, one could estimate an error correction model (ECM) if the variables were cointegrated and a stationary long run relationship existed between them.

Buehn and Schneider reexpress the MIMIC model as follows:

Structural part:

$$\eta_t = \gamma' x_t + \varsigma_t \qquad (42)$$

where $x_t' = (x_{1t}, x_{2t}, \dots, x_{qt})$ is a (1×q) vector of time series variable as indicated by the subscript $t$. Each time series $x_{it}, I = 1, \dots, q$ is a potenatial cause of the latent variable $\eta_t$. $\gamma' = (\gamma_1, \gamma_2, \dots, \gamma_q)$, a (1×q) vector of coefficients in the structural model describing the »causal« relationships between the latent variable and its causes.

Measurement part:

$$y_t = \lambda\eta_t + \varepsilon_t \qquad (43)$$

where $y_t' = (y_{1t}, y_{2t}, \dots, y_{pt})$ is a (1×p) vector of individual time series variables $y_{jt}, j = 1, \dots, p$. $\varepsilon_t = (\varepsilon_{1t}, \varepsilon_{2t}, \dots, \varepsilon_{pt})$ is a (p×1) vector of distrurbances where every $\varepsilon_{jt}, j = 1, \dots, p$ is a white noise term. Their (p×p) covariance matrix is given by $\Theta_\varepsilon$. The single $\lambda_j, j = 1, \dots, p$ in the (p×1) vector of regression coefficients $\lambda$, represents the magnitude of the expected change of the respective indicator for a unit change in the latent variable.

As before, reduced form equation is:

$$y_t = \Pi x_t + z_t \qquad (44)$$

where $\Pi = \lambda \gamma'$ and $z_t = \lambda \zeta_t + \varepsilon_t$. The error term $z_t$ in equation above is a (p×1) vector of linear combinations of the white noise error terms $\varsigma_t$ and $\varepsilon_t$ from the structural and measurement model, i.e. $z_t \sim (0, \Omega)$. The covariance matrix $\Omega$ is given as $Cov(z_t) = \lambda \lambda' \psi + \Theta_\varepsilon$.

We will denote the variables in the above model which are I(1) as $x_t$ and those that are I(0) as $v_t$. Above equation then becomes:

$$y_t = \Pi x_t + T v_t + z_t \qquad (45)$$

where $T = \lambda \beta'$ and $\tau' = (\tau_1, \tau_2, \ldots, \tau_r)$ is the (1×r) vector of coefficients of the I(0) variables in the structural relationship.

Every cointegration relationship has an error correction mechanism where the long run relationship leads to equilibrium and the short run relationship contains a dynamic mechanism (Engle and Granger, 1987). Thus, above equation can be written as:

$$\Delta y_t = A \Delta x_t + T v_t + K z_{t-1} + w_t \qquad (46)$$

where $\Delta y_t = y_t - y_{t-1}, \Delta x_t = x_t - x_{t-1}, z_{t-1} = y_{t-1} - \Pi x_{t-1}$ and $A$, $B$ and $K$ are coefficient matrices in this dynamic, short run model spefication. Furthermore, in this specification $A = \lambda \alpha'$ is the [p×(q-r)] coefficient matrix of the first differences of the I(1) causes, and $B = \lambda \beta'$ is the (p×r) coefficient matrix of the I(0) causes. The matrix $K = \lambda \kappa'$ is the (p×p) coefficient matrix for the long run disequilibrium's error correction term and $w_t \sim (0, \Omega)$ is a white noise disturbance. Together, Together, equations above define the EMIMIC model.

We now translate the model into our 2SLS framework and Jöreskog's analysis.

The variance-covariance matrix for the model in (45) can be written as below:

$$\Sigma = \begin{pmatrix} Var(x_t) & & \\ Cov(x_t, v_t) & Var(v_t) & \\ Cov(x_t, y_t) & Cov(v_t, y_t) & Var(y_t) \end{pmatrix} \qquad (47)$$

and in the notation above as:

$$\Sigma = \begin{pmatrix} \Phi_1^* & N & (\Phi_1^* \gamma^* + N\tau)\lambda' \\ N' & \Phi_2 & (N'\gamma^* + \Phi_2 \tau)\lambda' \\ \lambda(\gamma^{*\prime}\Phi_1^* + \tau'N') & \lambda(\gamma^{*\prime}N + \tau'\Phi_2) & \lambda(\gamma^{*\prime}\Phi_1^*\gamma^* + 2\gamma^{*\prime}N\tau + \tau'\Phi_2\tau)\lambda' + \theta^2 \end{pmatrix} \qquad (48)$$

This can be translated into Jöreskog's analysis and formula (14) as:

$$\Phi = \begin{pmatrix} \Phi_{1(q-r)\times(q-r)}^* & N_{(q-r)\times r} \\ N'_{r\times(q-r)} & \Phi_{2 r\times r} \end{pmatrix}$$

$$B = \begin{pmatrix} I_{q\times(q-r)} & 0_{q\times1} \\ 0_{p\times(q-r)} & \lambda_{p\times1} \end{pmatrix}, \Lambda = \begin{pmatrix} I_{(q-r)\times(q-r)} & I_{r\times r} \\ \gamma^{*\prime}_{1\times(q-r)} & \tau'_{1\times r} \end{pmatrix},$$

$$\Psi = \begin{pmatrix} 0_{(q-r)\times(q-r)} & 0_{(q-r)\times1} \\ 0_{1\times(q-r)} & I_{1\times1} \end{pmatrix}, \Theta = \begin{pmatrix} 0_{q\times q} & 0_{q\times p} \\ 0_{p\times q} & \Theta_{p\times p} \end{pmatrix} \quad (49)$$

The variance-covariance matrix for the model in (49) can be written as:

$$\Sigma = \begin{pmatrix} Var(\Delta x_t) & & & \\ Cov(\Delta x_t, v_t) & Var(v_t) & & \\ Cov(\Delta x_t, z_{t-1}) & Cov(v_t, z_{t-1}) & Var(z_{t-1}) & \\ Cov(\Delta x_t, \Delta y_t) & Cov(v_t, \Delta y_t) & Cov(z_{t-1}, \Delta y_t) & Var(\Delta y_t) \end{pmatrix} \quad (50)$$

and in the notation above as:

$$\Sigma = \begin{pmatrix} \Phi_3^* & M^{*\prime} & 0 & (\Phi_3^*\alpha_\Delta^* + M^{*\prime}\beta_\Delta)\lambda' \\ M^* & \Phi_2 & 0 & (M^*\alpha_\Delta^* + \Phi_2\beta_\Delta)\lambda' \\ 0 & 0 & \Omega^* & \lambda\kappa^{*\prime}\Omega^* \\ \lambda(\alpha_\Delta^{*\prime}\Phi_3^* + \beta_\Delta'M^*) & \lambda(\alpha_\Delta^{*\prime}M^{*\prime} + \beta_\Delta'\Phi_2) & \Omega^*\kappa^*\lambda' & \lambda(\alpha_\Delta^{*\prime}\Phi_3^*\alpha_\Delta^* + 2\alpha_\Delta^{*\prime}M^{*\prime}\beta_\Delta + \beta_\Delta'\Phi_2\beta_\Delta + \kappa^{*\prime}\Omega^*\kappa^*)\lambda' + \psi\lambda\lambda' + \Theta^2 \end{pmatrix} \quad (51)$$

This can be translated into Jöreskog's analysis and formula (14) as:

$$\Phi = \begin{pmatrix} \Phi_3^* & M^{*\prime} & 0 \\ M^* & \Phi_2 & 0 \\ 0 & 0 & \Omega^* \end{pmatrix}$$

$$B = \begin{pmatrix} I & 0 \\ 0 & \lambda \end{pmatrix}, \Lambda = \begin{pmatrix} I & I \\ \gamma^{*\prime} & \tau' \end{pmatrix},$$

$$\Psi = \begin{pmatrix} 0 & 0 \\ 0 & I \end{pmatrix}, \Theta = \begin{pmatrix} 0 & 0 \\ 0 & \Theta \end{pmatrix} \quad (52)$$

Below we present results of simulation studies of the performance of the above new estimators. Our simulation results are based on 10000 simulated data sets and corresponding 1000 resamples. We present results for three scenarios: 1) with short time series (t=20) and with only one I(1) variable; 2) with short time series (t=20) and several (three) I(1) variables; 3) with longer time series (t=100) and several (three) I(1) variables. We simulate three criteria of the MIMIC models: root mean square error of approximation (RMSEA), standardized root mean square residual (SRMR) and comparative fit index (CFI).

Results of simulation study for scenario 1, presented in Table 1, confirm the consistency of the proposed approaches. As demonstrated the 2SLS-MIMIC procedure leads to consistent and asymptotically normal estimator which is shown in Table 1 for both error criteria – with enlarging the sample size, the error of the estimates outperforms other, noninstrumented estimators by a sizable amount. Also, the fit of the model is significantly improved for both 2SLS-MIMIC as well as 2SLS-EMIMIC procedures.

**Table 1:** Simulation study, scenario 1

| RMSEA | MIMIC | DMIMIC | EMIMIC | 2SLS-MIMIC | 2SLS-EMIMIC |
|---|---|---|---|---|---|
| 50 | 0.1318 | 0.1261 | 0.1198 | 0.1102 | 0.1051 |
| 100 | 0.1081 | 0.0883 | 0.0910 | 0.0937 | 0.0799 |
| 200 | 0.0940 | 0.0724 | 0.0637 | 0.0759 | 0.0663 |
| 500 | 0.0733 | 0.0622 | 0.0510 | 0.0660 | 0.0477 |
| 1000 | 0.0587 | 0.0498 | 0.0418 | 0.0482 | 0.0344 |
| 2000 | 0.0458 | 0.0398 | 0.0368 | 0.0366 | 0.0275 |
| 5000 | 0.0389 | 0.0299 | 0.0291 | 0.0278 | 0.0231 |
| 10000 | 0.0296 | 0.0248 | 0.0224 | 0.0223 | 0.0201 |

| SRMR | MIMIC | DMIMIC | EMIMIC | 2SLS-MIMIC | 2SLS-EMIMIC |
|---|---|---|---|---|---|
| 50 | 0.1702 | 0.1691 | 0.1678 | 0.1622 | 0.1569 |
| 100 | 0.1413 | 0.1522 | 0.1493 | 0.1330 | 0.1177 |
| 200 | 0.1257 | 0.1111 | 0.1225 | 0.0931 | 0.0871 |
| 500 | 0.1044 | 0.0867 | 0.0882 | 0.0661 | 0.0775 |
| 1000 | 0.0793 | 0.0624 | 0.0697 | 0.0588 | 0.0581 |
| 2000 | 0.0682 | 0.0499 | 0.0522 | 0.0518 | 0.0407 |
| 5000 | 0.0566 | 0.0444 | 0.0402 | 0.0456 | 0.0297 |
| 10000 | 0.0430 | 0.0386 | 0.0350 | 0.0364 | 0.0226 |

| CFI | MIMIC | DMIMIC | EMIMIC | 2SLS-MIMIC | 2SLS-EMIMIC |
|---|---|---|---|---|---|
| 50 | 0.8611 | 0.8720 | 0.8801 | 0.9102 | 0.9119 |
| 100 | 0.8783 | 0.8894 | 0.8889 | 0.9193 | 0.9301 |
| 200 | 0.8871 | 0.8983 | 0.8978 | 0.9377 | 0.9394 |
| 500 | 0.9048 | 0.9073 | 0.9157 | 0.9564 | 0.9488 |
| 1000 | 0.9229 | 0.9255 | 0.9341 | 0.9660 | 0.9678 |
| 2000 | 0.9322 | 0.9440 | 0.9434 | 0.9757 | 0.9775 |
| 5000 | 0.9415 | 0.9629 | 0.9528 | 0.9854 | 0.9873 |
| 10000 | 0.9509 | 0.9725 | 0.9624 | 0.9953 | 0.9971 |

Source: Own calculations.

With more I(1) variables in the dynamic version of the (correct) model (Table 2, i.e. scenario 2), the difference in performance between 2SLS-MIMIC and 2SLS-EMIMIC procedures grows in favor of the latter, which is particularly pronounced in the fit of the model, where both models significantly outperform the noninstrumented ones.

**Table 2:** Simulation study, scenario 2

| RMSEA | MIMIC | DMIMIC | EMIMIC | 2SLS-MIMIC | 2SLS-EMIMIC |
|---|---|---|---|---|---|
| 50 | 0.1437 | 0.1223 | 0.1330 | 0.1256 | 0.0914 |
| 100 | 0.1189 | 0.0812 | 0.1083 | 0.0843 | 0.0863 |
| 200 | 0.1110 | 0.0622 | 0.0714 | 0.0819 | 0.0696 |
| 500 | 0.0638 | 0.0566 | 0.0597 | 0.0733 | 0.0387 |
| 1000 | 0.0516 | 0.0513 | 0.0376 | 0.0511 | 0.0388 |
| 2000 | 0.0366 | 0.0410 | 0.0338 | 0.0311 | 0.0316 |
| 5000 | 0.0373 | 0.0344 | 0.0340 | 0.0242 | 0.0263 |
| 10000 | 0.0278 | 0.0295 | 0.0219 | 0.0216 | 0.0193 |

| SRMR | MIMIC | DMIMIC | EMIMIC | 2SLS-MIMIC | 2SLS-EMIMIC |
|---|---|---|---|---|---|
| 50 | 0.1702 | 0.1725 | 0.2014 | 0.1654 | 0.1757 |
| 100 | 0.1653 | 0.1750 | 0.1478 | 0.1476 | 0.1377 |
| 200 | 0.1220 | 0.1222 | 0.1286 | 0.0773 | 0.0836 |
| 500 | 0.1127 | 0.0832 | 0.0749 | 0.0787 | 0.0853 |
| 1000 | 0.0706 | 0.0736 | 0.0676 | 0.0624 | 0.0651 |
| 2000 | 0.0778 | 0.0524 | 0.0481 | 0.0518 | 0.0415 |
| 5000 | 0.0521 | 0.0413 | 0.0374 | 0.0542 | 0.0267 |
| 10000 | 0.0430 | 0.0348 | 0.0304 | 0.0419 | 0.0181 |

| CFI | MIMIC | DMIMIC | EMIMIC | 2SLS-MIMIC | 2SLS-EMIMIC |
|---|---|---|---|---|---|
| 50 | 0.8183 | 0.8233 | 0.8097 | 0.8219 | 0.8496 |
| 100 | 0.8271 | 0.8319 | 0.8259 | 0.8377 | 0.8552 |
| 200 | 0.8360 | 0.8406 | 0.8424 | 0.8439 | 0.8711 |
| 500 | 0.8449 | 0.8594 | 0.8508 | 0.8521 | 0.8992 |
| 1000 | 0.8630 | 0.8683 | 0.8593 | 0.8687 | 0.9056 |
| 2000 | 0.8723 | 0.8763 | 0.8679 | 0.8857 | 0.9223 |
| 5000 | 0.8909 | 0.8846 | 0.8853 | 0.9044 | 0.9308 |
| 10000 | 0.9004 | 0.9033 | 0.8941 | 0.9231 | 0.9494 |

Source: Own calculations.

Finally, with longer time series, the performance of the 2SLS-EMIMIC as compared to 2SLS-MIMIC is not so good anymore. While it outperforms the latter in terms of error criteria, it lags behind in fit index which might show some problems with cointegration properties for longer time series of the proposed 2SLS-EMIMIC estimation.

**Table 3:** Simulation study, scenario 3

| RMSEA | MIMIC | DMIMIC | EMIMIC | 2SLS-MIMIC | 2SLS-EMIMIC |
|---|---|---|---|---|---|
| 50 | 0.1595 | 0.1431 | 0.1543 | 0.1281 | 0.0796 |
| 100 | 0.1296 | 0.0747 | 0.0932 | 0.0792 | 0.0992 |
| 200 | 0.1143 | 0.0697 | 0.0621 | 0.0967 | 0.0745 |
| 500 | 0.0549 | 0.0583 | 0.0614 | 0.0586 | 0.0452 |
| 1000 | 0.0573 | 0.0415 | 0.0335 | 0.0577 | 0.0350 |
| 2000 | 0.0297 | 0.0349 | 0.0288 | 0.0352 | 0.0281 |
| 5000 | 0.0366 | 0.0412 | 0.0289 | 0.0206 | 0.0308 |
| 10000 | 0.0300 | 0.0242 | 0.0257 | 0.0255 | 0.0154 |

| SRMR | MIMIC | DMIMIC | EMIMIC | 2SLS-MIMIC | 2SLS-EMIMIC |
|---|---|---|---|---|---|
| 50 | 0.1821 | 0.2001 | 0.2416 | 0.1820 | 0.1968 |
| 100 | 0.1620 | 0.1400 | 0.1198 | 0.1727 | 0.1391 |
| 200 | 0.1073 | 0.1259 | 0.1337 | 0.0618 | 0.0786 |
| 500 | 0.0969 | 0.0923 | 0.0630 | 0.0661 | 0.0963 |
| 1000 | 0.0826 | 0.0810 | 0.0757 | 0.0630 | 0.0775 |
| 2000 | 0.0762 | 0.0529 | 0.0457 | 0.0528 | 0.0340 |
| 5000 | 0.0516 | 0.0331 | 0.0441 | 0.0629 | 0.0251 |
| 10000 | 0.0495 | 0.0306 | 0.0286 | 0.0335 | 0.0197 |

| CFI | MIMIC | DMIMIC | EMIMIC | 2SLS-MIMIC | 2SLS-EMIMIC |
|---|---|---|---|---|---|
| 50 | 0.8168 | 0.8392 | 0.8178 | 0.8869 | 0.8836 |
| 100 | 0.8250 | 0.8470 | 0.8260 | 0.8958 | 0.8925 |
| 200 | 0.8333 | 0.8551 | 0.8342 | 0.9137 | 0.9014 |
| 500 | 0.8416 | 0.8736 | 0.8509 | 0.9228 | 0.9194 |
| 1000 | 0.8584 | 0.8830 | 0.8679 | 0.9413 | 0.9286 |
| 2000 | 0.8670 | 0.8926 | 0.8766 | 0.9601 | 0.9379 |
| 5000 | 0.8757 | 0.9022 | 0.8941 | 0.9697 | 0.9567 |
| 10000 | 0.8844 | 0.9116 | 0.9120 | 0.9891 | 0.9662 |

Source: Own calculations.

### 4. Application – precarious work of the elderly

As mentioned below, the data was sourced from SHARE, an EU-wide panel study of more than 120.000 individuals aged 50+, currently covering 27 countries and Israel. So far, six waves of the study have been conducted: 2004, 2006-07, 2008-09, 2010-11, 2013, 2015 and 2017, with data available for the first six. Information is collected by means of Computer-Assisted Personal Interview (CAPI) questionnaires, physical measurements and fill-in questionnaires across several modules containing demographic, health, psychological and socio-economic information (see Börsch-Supan et al., 2013). The 2013 wave (SHARELIFE) adopted a retrospective perspective and recorded life histories.

Both waves 3 and 6 of the study were considered as sources for the construction of the index, wave 6 being the most recent and comprehensive, and wave 3 being information-rich (i.e. panel data). In order to define the potential population, question ep005_CurrentJobSit1, which describes the respondents' current job situation, was used. One relevant category ("employed or self-employed") was identified. The distribution of respondents from waves 3 and 6 across it is provided in the table below.

Given the fact that information from both waves 3 and 6 was available for only 1048 individuals, we opted to include only wave 6 in the production of the index, as it comprises 16716 individuals. Once the complete cases were identified in function of the chosen variables (cf. infra), 5594 observations were retained.

The operationalization started by translating Kalleberg's (2009) definition of precarious work as "uncertain, unpredictable and risky" into several dimensions, i.e. employment stability, income and working conditions.

In order to define those dimensions, we took two steps. We first looked at the existing indices summarized in Table 3 above, and at the same time we pre-selected all the questions dealing with the working life of the SHARE respondents from wave 6 (module EP, for a description of the data see below). 423 questions were selected, among which several instrumental questions (e.g. bracket values used to estimate income amounts) and looped questions (e.g. relating to each of the respondent's income sources or former jobs). Then only the questions were retained that related to the respondents' present job. Then the questions were labelled following the dimensions defined by Tangian (2007) and Olsthoorn (2014) and the best-fitting approach was selected and amended.

Olsthoorn was selected because the logics on which his operationalization is built take into account the substitutability-additionality continuum (cf. supra). Tangian's dimensions were selected because of three main reasons. First, his dimensions and sub-dimensions cover a broad spectrum of the employment relationship as opposed to, for instance, Cranford and colleagues (2003), who focus on regulatory aspects. Second, his dimensions are far enough from the indicator level (i.e. abstract enough) as to be applied to different data. Conversely, other indices such as Clark (2005) and Böckermann (2004) are very closely tailored to the data collection. Other, such as Greenhalgh and Rosenblatt (1984) and Loughlin and Murray (2013) are difficult to operationalize by proxy. Third, he uses European surveys in order to construct his indicators, as is the case in the present analysis.

In practice, the coding largely overlapped: Olsthoorn's deprivation dimension was similar to Tangian's income, and insecurity and employment instability were more or less the same as well. Tangian's "employability" has however no counterpart in Olsthoorn's classification.

In a second step, the subdimensions were amended on the basis of the available data. A new dimension ("subjective appreciation") was introduced, which echoes the subjective perspective of several of the indices mentioned above. The final result included thus five dimensions: income, stability, employability, integration in social security, and subjective appreciation.

The definition of the dimensions and subdimensions was finished with the definition of the causal relationships between the dimensions which, as it has been said above, has not been problematized by the literature.

---

1 As opposed to Srakar & Prevolnik Rupel (2017) question ep002, which identifies those respondents who work in spite of self-identifying as non-working in ep005, is not taken into consideration when defining the population, due to the fact that several questions relating to the working relationship are only posed to those respondents self-identifying as working in ep005.

A first step in defining the (potential) causal relationships between the dimensions was to identify the analytical level at which they are located. Income, integration in social security and employment stability are located at the level of the work, and some of their features may arguably be located at the institutional regime. Subjective appreciation, on the other hand, as well as employability, are located at the analytical level of the worker. The possible correlations between the different dimensions or levels have not been hypothesized, and those authors who incorporate both objective and subjective dimensions into their indices ((Tangian, 2007; Vives et al., 2010) do not elaborate on the possible relations between them, and regard them both simply as components of precarious work. This overlooks two possible causal relationships:

1. The subjective appreciation of the worker is influenced by the objective components at the job level: workers with unstable employment, a low income and a low degree of integration in social security are more likely to be overall less satisfied with their jobs than workers in a less precarious situation.
2. Employability is both a function and a determinant of the objective dimensions: a worker with scarce training and career advancement opportunities is more likely to have low stability, low income and a low degree of integration in social security, and jobs with those characteristics are more likely to foster low employability than otherwise.

Whereas the case can be made for subjective appreciation as being a part of precarious work as such, it is difficult to ignore the causal links hypothesized above. Therefore, the causal model used for the index incorporates them both. In this model, precarious work is regarded as a latent variable defined by the three objective dimensions and employability, which in turn affects the subjective appreciation of the worker. In other words, subjective appreciation does not strictly belong to the "precarious work" construct. The result of the exercise is portrayed in the figure below.

**Figure 1:** The hypothesized relations between the dimensions of the precarious work concept

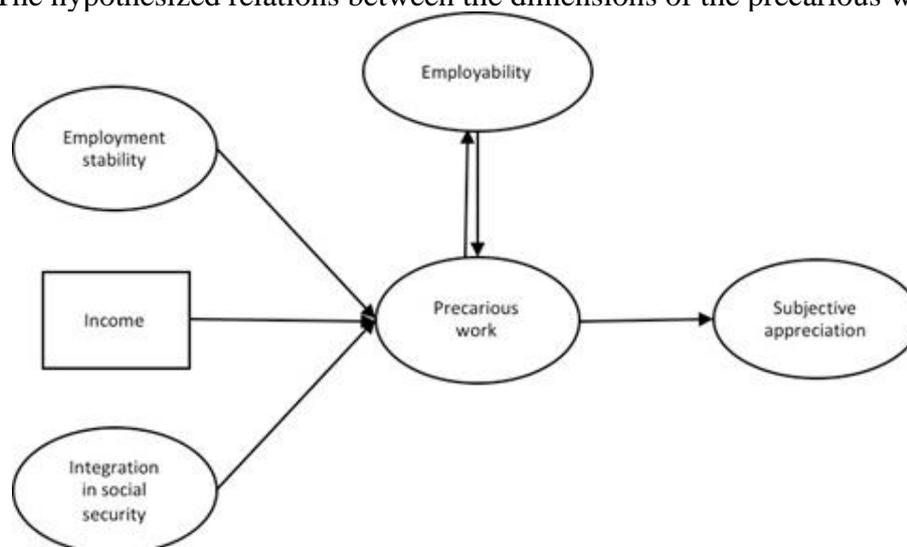

Source: Own elaboration.

Once the indicators were all labelled, those related to the respondents' current job were selected. All variables relating to either his whole career or his former job were dropped. In addition, missing values were assessed. An arbitrary cutoff value was set at 60%, due mainly to looped variables. Questions where information was missing for more than 50% of the population, and which were not part of any specific if-loops, were dismissed. Variables related to loops in main variables (such as in the case of persons having more than one job), or lists (such as income sources) for which missing values are higher than 60% were included, as they relate to specific sub-populations within the main group. Questions asked specifically to the 4929 first-time respondents (i.e. "baseline" vs. "longitudinal" questionnaire, for which variable mn101 = 0) were however not included in the index.

The final selection includes 25 variables constructed from the Employment and Pensions module of the survey and one variable (income) imported from the Generated Variables module. One looped variable (i.e. pensions) was summarized in a single indicator. All variables were recoded so they would point in the same direction, i.e. low values coincide with low precarity, high values coincide with high precarity.

The selected indicators, as recoded for the analysis, are described in Table 5 below. Table 6 provides some basic statistics.

**Table 4:** List of used variables

| Variable name | Description | Dimension | Type |
|---|---|---|---|
| income_decile | Income deciles | Income | Discrete |
| job_number | Number of jobs | Employment stability | Dichotomous |
| job_self | Self employment | Employment stability | Dichotomous |
| job_satisfaction | Satisfied with (main) job | Subjective appreciation | Ordinal |
| job_phdeman | (Main) job physically demanding | Subjective appreciation | Ordinal |
| job_timepress | Time pressure due to a heavy workload in (main) job | Subjective appreciation | Ordinal |
| job_freedom | Little freedom to decide how I do my work in (main) job | Subjective appreciation | Ordinal |
| job_newskills | Opportunity to develop new skills in (main) job | Employability | Ordinal |
| job_support | Receive support in difficult situations in (main) job | Subjective appreciation | Ordinal |
| job_recognition | Receive recognition for work in (main) job | Subjective appreciation | Ordinal |
| job_sat_salary | Salary or earnings are adequate in (main) job | Subjective appreciation | Ordinal |
| job_promotion | Poor prospects for (main) job advancement | Employability | Ordinal |
| job_security | Poor (main) job security | Subjective appreciation | Ordinal |
| pensions_future | Number of pensions claimed in the future | Integration in social security | Discrete |
| pensions_benefit | Amount of benefits in the future | Integration in social security | Continuous |
| Tenure | Number of years in current job as percentage of age | Employment stability | Continuous |
| job_term | Type of contract | Employment stability | Ordinal |
| pensions_years_ratio | Number of years the respondent has been contributing to pensions as a ratio of tenure | Integration in social security | Continuous |
| pensions_comp_ratio | compulsory pensions as a ratio of the total number of of pensions | Integration in social security | Continuous |
| Job_hours | Total hours usually working per week | Employment stability | Continuous |

Source: SHARE Wave 6, own modifications.

**Table 5:** Descriptive statistics of the used variables

| Variable name | Median | Mean | Std dev | Minimum | Maximum |
|---|---|---|---|---|---|
| income_decile | 40000.00 | 47311.00 | 26405.00 | 10000.00 | 100000.00 |
| job_number | 0.00 | 0.07 | 0.26 | 0.00 | 10000.00 |
| job_self | 0.00 | 0.03 | 0.18 | 0.00 | 10000.00 |
| pensions_future | -10000.00 | -11113.00 | 0.83 | -50000.00 | 0.00 |
| pensions_benefit | 8800000.00 | 7089680.00 | 4938989.00 | 0.00 | 11889230.00 |
| tenure | 0.66 | 0.65 | 0.22 | 0.20 | 10000.00 |
| job_term | 0.00 | 0.09 | 0.28 | 0.00 | 10000.00 |
| pensions_years | -360000.00 | -290637.00 | 177777.00 | -460000.00 | 0.00 |
| pensions_comp_ratio | -10000.00 | -0.68 | 0.45 | -10000.00 | 0.00 |
| job_hours | 400000.00 | 367460.00 | 60987.00 | 250000.00 | 430000.00 |
| job_satisfaction | 20000.00 | 23349.00 | 0.67 | 0.00 | 30000.00 |
| job_phdeman | 10000.00 | 13845.00 | 10578.00 | 0.00 | 30000.00 |
| job_timepress | 10000.00 | 14478.00 | 0.92 | 0.00 | 30000.00 |
| job_freedom | 10000.00 | 11547.00 | 0.94 | 0.00 | 30000.00 |
| job_newskills | 20000.00 | 18063.00 | 0.88 | 0.00 | 30000.00 |
| job_support | 20000.00 | 19261.00 | 0.81 | 0.00 | 30000.00 |
| job_recognition | 20000.00 | 18388.00 | 0.83 | 0.00 | 30000.00 |
| job_sat_salary | 20000.00 | 15995.00 | 0.87 | 0.00 | 30000.00 |
| job_promotion | 20000.00 | 19854.00 | 0.92 | 0.00 | 60000.00 |
| job_security | 10000.00 | 0.89 | 0.89 | 0.00 | 30000.00 |

Source: SHARE Wave 6, own modifications.

The calculation of the index's values was conducted, following the recommendation in OECD (2008), using regression techniques, *in casu* Multiple Indicator Multiple Cause (MIMIC) analysis, which is part of the larger family of Structural Equation Modelling (SEM). It was conducted using library lavaan from the R software package ("The lavaan Project," n.d.), and comprised three steps: the definition of precarious work as a latent variable in a MIMIC model, the introduction of an instrumental variable to correct for the causal loop of employability and precarious work, and the calculation of the index on the basis of the predicted values of the latent variable. The three steps are set out below.

The graphic representation of the model following the MIMIC conventions is depicted in the next page. It is clear from the figure that the model corresponds largely to the causal model depicted in Figure 1. However, some adaptations require clarifications. First, the employability dimension (portrayed as a latent variable in the model above) has been replaced by a proxy, Job_new_skills. Second, the regression coefficients from and to employability were set to be equal in order to obtain an identified model (otherwise the software is unable to estimate the standard error terms, cf. infra). Third, some of the variables belonging to employment stability (tenure, job_number and job_hours) are regarded as reflective rather than formative indicators following our causal model: the number of hours someone works, the number of years they stay in their job and the number of jobs they do besides their first job do not *cause* precarious work, but are rather hypothesized to be the effect of the employment stability linked to the formal status of the worker.

**Figure 2:** MIMIC model for precarious work

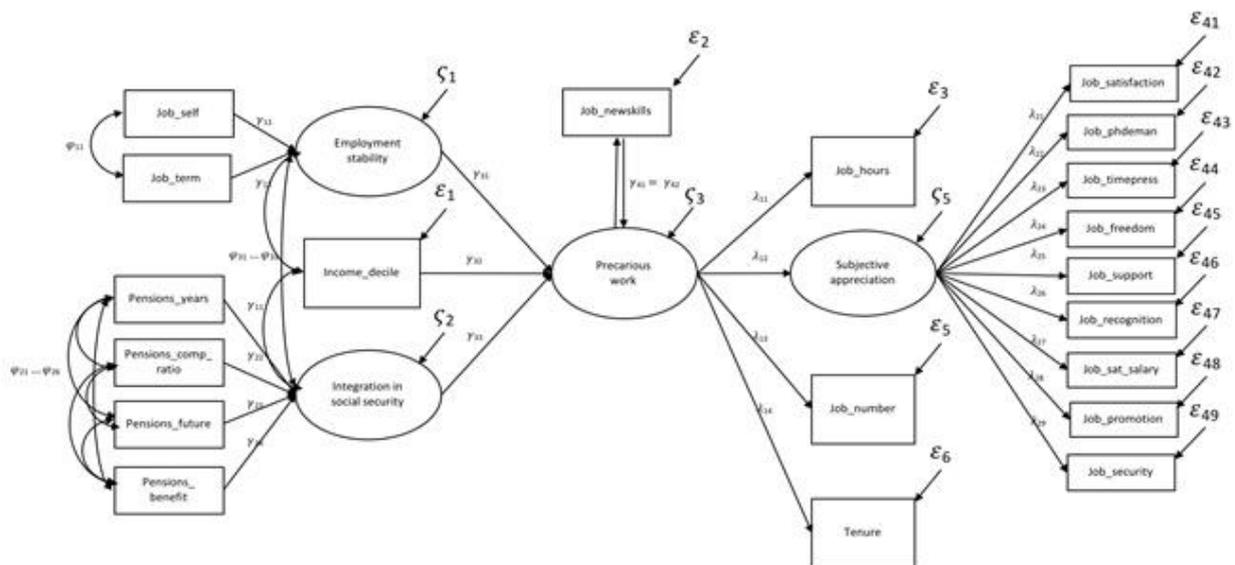

Source: Own elaboration.

Since the lavaan package does not include any built-in mechanism to incorporate instrumental variables to MIMIC, the IV for employability was introduced by regressing the variable in question on some predictors exogenous to the model and then replacing the observed values with the estimates from the regression in the MIMIC model, making the analysis a de facto two-stage regression. In our case, job_new_skills was regressed on a number of demographic and personal characteristics from the respondents (age, gender, years of education, health status, self-reported writing abilities, and results on a numeracy test), as well as country dummies. The results of the regression are reported in the following table.

**Table 6:** Linear regression of employability

| Variable name | Coef. | Sig. |
|---|---|---|
| Intercept | 0.30 | *** |
| gender | 0.06 | ** |
| yedu | -0.03 | *** |
| sphus | 0.10 | *** |
| writing | 0.06 | *** |
| numeracy2 | -0.03 | ** |
| Germany | -0.03 | |
| Sweden | -0.15 | ** |
| Spain | 0.08 | * |
| Italy | 0.28 | *** |
| France | 0.26 | *** |
| Denmark | -0.18 | ** |
| Greece | 0.36 | *** |
| Switzerland | -0.23 | *** |
| Belgium | 0.08 | |
| Israel | 0.08 | *** |
| CzechRepublic | -0.04 | |
| Poland | 0.23 | *** |
| Luxemburg | 0.01 | |
| Portugal | -0.12 | |
| Slovenia | -0.03 | |
| Estonia | -0.20 | |
| *Nr. observations* | 5560 | |
| *Mult. R-squared* | 0.082 | |
| *Adj. R-squared* | 0.081 | |
| *F-Statistic* | 52.45 | |
| *df* | 21 and 12313 | |
| *p-value* | 0.00 | |

Note: Significance: * <0.1 ** < 0.01 *** <0.001.
Source: SHARE, Wave 6 (own modifications).

As it can be seen from the table, our proxy for employability is significantly influenced by a few predictors: gender (female tend to be less employable than men), health status (bad health reduces the opportunities to develop new skills), education (the higher the number of years and education, the lower the employability), writing (the poorer the ability, the lower the employability) and numeracy (same relation as writing). Among the demographic and personal characteristics, the largest influence comes from health status. However, the largest influence comes from the country dummies: workers in Italy, France, Greece, Israel and Poland tend to fare significantly worse regarding employability, and those in Denmark and Switzerland better.

The third step in the construction of the index was the MIMIC analysis, from which the index values were obtained. The results from the analysis are summarized in the following table. Results are provided for two models, with and without the instrumental variable-correction.

**Table 7:** Results of the MIMIC regression

| | Model 1 | | Model 2 | |
|---|---|---|---|---|
| | Coef. (standardized) | Sig. | Coef. (standardized) | Sig. |
| *Latent variables* | | | | |
| **Precar** | | | | |
| Pensions | 1.00 | | 1.00 | |
| Income | 5061.00 | *** | 3039.00 | *** |
| Job_newskills | 0.67 | *** | | *** |
| Est_job_skills | | | 0.47 | |
| Stability | 0.30 | *** | 0.13 | *** |
| **Subjective** | | | | |
| job_satisfaction | 1.00 | | 1.00 | |
| job_phdeman | 0,738 | *** | 0,758 | *** |
| job_timepress | 0,724 | *** | 0,741 | *** |
| job_freedom | 0,901 | *** | 0,897 | *** |
| job_support | 1220.00 | *** | 1216.00 | *** |
| job_recognition | 1496.00 | *** | 1502.00 | *** |
| job_sat_salary | 1160.00 | *** | 1168.00 | *** |
| job_promotion | 0.52 | *** | -0.51 | *** |
| job_security | 0.63 | *** | 0.62 | *** |
| **Pensions** | | | | |
| Pensions_future | 1.00 | | 1.00 | |
| Pensions_compulsory | -0.84 | *** | -0.83 | *** |
| Pensions_years | 18970.00 | *** | 18785.00 | *** |
| **Stability** | | | | |
| Job_term | 1.00 | | 1.00 | |
| Job_self | 0.09 | * | 0.08 | |
| *Regressions* | | | | |
| Job_newskills – precar | 0.00 | | 0.00 | |
| Subj – precar | 0.64 | *** | 0.41 | *** |
| Tenure - precar | 0.24 | *** | -0.01 | |
| Job_number - | -0.11 | *** | -0.12 | *** |
| *Goodness of fit tests* | | | | |
| Tucker-Lewis Index | 0.83 | | 0.83 | |
| Comparative Fit Index | 0.86 | | 0.85 | |
| RMSEA | 0.07 | *** | 0.07 | *** |
| Model fit test statistic | 3262.38 | *** | 3547.31 | *** |

Note: Significance: * <0.1; ** < 0.01; *** <0.001.
Source: SHARE, Wave 6 (own modifications).

Several preliminary conclusions can be made from the model. First, if we look at the fit with strict standards, the model is mediocre at best and pointless at worst. It should be pointed, however, that the exploratory scope of the study allows some slack in terms of model accuracy. Second, the largest single predictor for precarious work seems to be income: a low income is associated with high precarity. Employability seems to exert some influence as well, though much smaller. Third, most of the effects go in the hypothesized direction (i.e. high values of the variable are associated with high values of precarious work) with some exceptions: the number of jobs (more precarity is associated with people who have only one job as opposed to people with two jobs) and the perception of promotion opportunities in the model with the employability estimates.

On the basis of the model above, the estimated scores for the variable "precar" were averaged by country, rescaled (0-1) and used as the precarious work index. The results can be seen in the table and graph below.

**Table 8:** Precarious work index by country

| Country | Precar – Model 1 | Precar-Model 2 |
|---|---|---|
| Austria | 0.44 | 0.25 |
| Belgium | 0.52 | 0.35 |
| Czech Republic | 0.42 | 0.43 |
| Denmark | 0.00 | 0.10 |
| Estonia | 0.47 | 0.76 |
| France | 0.70 | 0.43 |
| Germany | 0.41 | 0.33 |
| Greece | 1.00 | 1.00 |
| Israel | 0.55 | 0.68 |
| Italy | 0.89 | 0.79 |
| Luxembourg | 0.45 | 0.45 |
| Poland | 0.81 | 0.76 |
| Portugal | 0.58 | 0.50 |
| Slovenia | 0.51 | 0.48 |
| Spain | 0.75 | 0.78 |
| Sweden | 0.08 | 0.00 |
| Switzerland | 0.18 | 0.14 |
| **Total** | **0.52** | **0.49** |

Source: SHARE, Wave 6 (own modifications).

**Figure 3:** Precarious work index by country

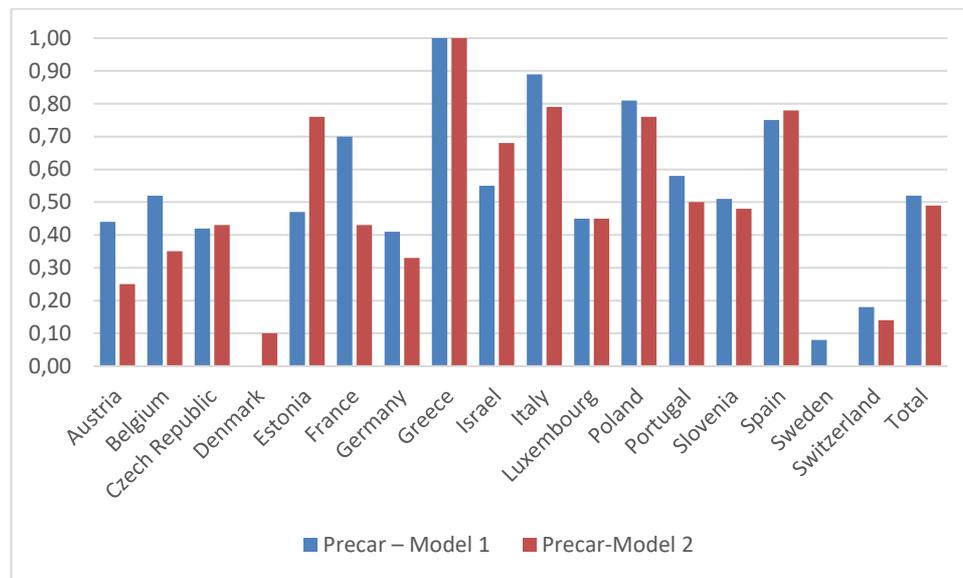

Source: SHARE, Wave 6 (own modifications).

The results from Model 1 seem, in general terms, to echo the institutional differences among regimes: the lowest levels of precarious work can be found in Scandinavia (Denmark and Sweden), with Switzerland following closely. The highest levels are, by contrast, to be found in Southern Europe (Spain, Greece and Italy). The continental welfare regimes from Western and Central Europe (Austria, Belgium France, Germany and Luxembourg) are somewhere in the middle, whereas the post-communist states show no common denominator: Slovenia and Estonia are closer to central Europe, whereas Poland displays larger scores.

The results from model 2 are largely similar, with two exceptions: Estonia displays a much lower degree of precariousness, as well as France.

5. **Concluding remarks**

The article presented is to our knowledge the second one (first one being Tekwe et al., 2014 – their article addressed endogeneity in MIMIC in a very limited context related specifically to measurement error problems) which explicitly models reverse causality in MIMIC models which can arise very frequently. We present a novel estimation procedure, based on Bollen's 2SLS estimator and transformation into Jöreskog's general covariance structure analysis. We are able to derive three new estimation procedures and show their consistency and asymptotic normality (for 2SLS-MIMIC). While the task remains (we are working on this presently) to derive the asymptotics also for the 2SLS-EMIMIC estimator, the results of simulation studies confirm the validity of the procedure and desired properties of the new estimators.

We have to mention several limitations of the study. Firstly, there exist significant critiques of the method of MIMIC for estimating shadow economy and other concepts. As has been shown by e.g. Breusch (2016), MIMIC is not always a proper modelling technique to estimate the latent concepts under question. Also, the IV context could be developed in more depth, related to overidentification issues and including estimators such as LIML, FIML, 3SLS and different types of GMM approaches. Furthermore, we address only maximum likelihood MIMIC estimation and do not relate to other two approaches at hand: econometric and factor analytic one (mentioned already in the original article of Jöreskog and Goldberger, 1975). For future work it would be important to address also those points more properly.